# Perspective on recent developments and challenges in regulatory and systems genomics

Julia Zeiltinger[1,2,#], Sushmita Roy[3,4,#], Ferhat Ay[5-7,#], Anthony Mathelier[8-10,#], Alejandra Medina-Rivera[11,#], Shaun Mahony[12,#], Saurabh Sinha[13,14,#], Jason Ernst[15-20,#]

[1] Stowers Institute for Medical Research, Kansas City, MO 64112, USA
[2] Department of Pathology & Laboratory Medicine, The University of Kansas Medical Center, Kansas City, KS 66160, USA
[3] Department of Biostatistics and Medical Informatics, University of Wisconsin-Madison, Madison, 53715, USA
[4] Wisconsin Institute for Discovery, University of Wisconsin-Madison, Madison, 53715, USA
[5] Centers for Autoimmunity, Inflammation and Cancer Immunotherapy, La Jolla Institute for Immunology, La Jolla, CA, USA.
[6] Bioinformatics and Systems Biology Program, University of California, San Diego, La Jolla, CA, USA.
[7] Department of Pediatrics, University of California, San Diego, La Jolla, CA, USA.
[8] Centre for Molecular Medicine Norway (NCMM), Nordic EMBL Partnership, University of Oslo, 0318 Oslo, Norway
[9] Department of Medical Genetics, Institute of Clinical Medicine, University of Oslo and Oslo University Hospital, Oslo, Norway
[10] Department of Pharmacy, University of Oslo, Oslo, Norway
[11] Laboratorio Internacional de Investigación sobre el Genoma Humano, Universidad Nacional Autónoma de México, Campus Juriquilla, Blvd. Juriquilla 3001, 76230, Santiago de Querétaro, México.
[12] Center for Eukaryotic Gene Regulation, Department of Biochemistry and Molecular Biology, The Pennsylvania State University, University Park, PA 16802, USA
[13] Walter H. Coulter Department of Biomedical Engineering, Georgia Institute of Technology, Atlanta, GA, USA.
[14] H. Milton Stewart School of Industrial and Systems Engineering, Georgia Institute of Technology, Atlanta, GA, USA.
[15] Department of Biological Chemistry, University of California, Los Angeles, Los Angeles, CA 90095, USA
[16] Computer Science Department, University of California, Los Angeles, Los Angeles, CA 90095, USA
[17] Department of Computational Medicine, University of California, Los Angeles, Los Angeles, CA 90095, USA
[18] Eli and Edythe Broad Center of Regenerative Medicine and Stem Cell Research at University of California, Los Angeles, Los Angeles, CA 90095, USA
[19] Jonsson Comprehensive Cancer Center, University of California, Los Angeles, Los Angeles, CA 90095, USA
[20] Molecular Biology Institute, University of California, Los Angeles, Los Angeles, CA 90095, USA

# Corresponding authors: jbz@stowers.org, sroy@biostat.wisc.edu, ferhatay@lji.org, anthony.mathelier@ncmm.uio.no, amedina@liigh.unam.mx, mahony@psu.edu, saurabh.sinha@bme.gatech.edu, jason.ernst@ucla.edu

# Abstract

Predicting how genetic variation affects phenotypic outcomes at the organismal, cellular, and molecular levels requires deciphering the cis-regulatory code, the sequence rules by which non-coding regions regulate genes. In this perspective, we discuss recent computational progress and challenges towards solving this fundamental problem. We describe how cis-regulatory elements are mapped and how their sequence rules can be learned and interpreted with sequence-to-function neural networks, with the goal of identifying genetic variants in human disease. We also discuss how studies of the 3D chromatin organization could help identifying long-range regulatory effects and how current methods for mapping gene regulatory networks could better describe biological processes. We point out current gaps in knowledge along with technical limitations and benchmarking challenges of computational methods. Finally, we discuss newly emerging technologies, such as spatial transcriptomics, and outline strategies for creating a more general model of the cis-regulatory code that is more broadly applicable across cell types and individuals.

# Glossary

- **Cis-regulatory element (CRE):** A DNA region in the genome bound by transcription factors or other proteins and contributes to gene regulation. It regulates a target gene in *cis,* thus its proximity to the gene on DNA is important. The most commonly studied CREs are enhancers, promoters, and architectural elements.
- **Cis-regulatory code:** The set of rules by which CREs are read by transcription factors and control gene expression.
- **Enhancer:** A CRE harboring motifs for one or more transcription factors with the ability to become active in one or more cellular conditions, resulting typically in the enhancement of transcription of a nearby gene. When not active, enhancers may be bound by repressive transcription factors.
- **Transcription factor motif:** Short DNA sequence pattern recognized by a transcription factor through protein-DNA interactions.
- **Position Weight Matrix (PWM):** a mathematical model for representing a transcription factor binding motif, where the frequencies of each base are summarized for each position.
- **Chromatin state:** The state in which genomic regions are found in vivo in the context of nucleosomes. Of specific interest are the chromatin states that change dynamically depending on the cellular conditions, such as DNA accessibility, histone modifications, and other bound proteins, since they are often the cause or effect of ongoing regulatory processes of transcriptional regulation.

- **ChIP-seq:** Chromatin immunoprecipitation sequencing. An assay for genome-wide profiling of transcription factor binding, histone modifications, and other features of chromatin state that can be specifically targeted by an antibody.
- **ATAC-seq:** Assay for transposase-accessible chromatin with sequencing, an experimental method for the genome-wide profiling of chromatin accessibility. scATAC-seq refers to the single-cell version.
- **scRNA-seq:** Genome-wide assay that measures RNA abundance in single cells.
- **Multi-omics assay:** Assays that measure multiple data modalities simultaneously, ideally as a single-cell assay. The most common example is the combination of RNA-seq and ATAC-seq.
- **Hi-C:** A high throughput technique for mapping the 3D structure of chromatin by mapping the pairwise contact frequencies between genomic regions.
- **Contact map:** A description of the 3D chromatin structure as mapped by Hi-C or related techniques, which is useful for understanding regulatory interactions among CREs and genes.
- **CTCF**: CCCTC-binding factor, a protein with a major role in regulating the 3D chromatin structure.
- **Loop extrusion:** A model that proposes that long-range cis-interactions within a DNA molecule are generated by loop extrusion factors (e.g., cohesin) that bind to DNA and reel flanking regions of the same DNA molecule into a loop that is demarcated by insulating elements (e.g., CTCF binding in convergent orientation).
- **Neural network:** A type of machine learning model, which can be trained to make accurate predictions from large amounts of complex data, typically by allowing many flexible parameters without specifying specific variables, features, or their relationships (black box model).
- **Deep learning model:** A neural network model with many layers (is "deep"), typically used to learn complex features of the cis-regulatory code.
- **Sequence-to-function model:** A neural network trained to predict experimental data ("function") from DNA sequences. The model architecture and sequence input length can vary depending on whether transcription factor binding, chromatin accessibility, histone modification data, gene expression or genome contact maps are predicted. Examples are convolutional neural networks and transformers.
- **Interpreting a neural network/deep learning model:** Using specific interpretation tools to open a black-box model to understand what features and rules a neural network model learned during training.
- **Single-nucleotide polymorphisms (SNPs):** Genomic sequences in which specific bases (A, C, T, or G) differ between individuals.
- **Genome-wide association studies (GWAS):** Statistical analysis of how genetic variants (usually SNPs) in individuals are associated with traits or diseases.
- **Gene regulatory network (GRN):** A collection of direct regulatory relationships between transcription factors, CREs, and target genes, often used as a model for expression changes between cellular conditions.

- **cis-GRN:** GRNs reconstructed based on analyzing cis-regulatory elements and TF motifs.
- **trans-GRN:** GRNs reconstructed based on analyzing co-expression between TFs and target genes.

## The fundamental problem of the cis-regulatory code

Predicting how genetic variation affects phenotypic outcomes at the organismal, cellular, and molecular levels is a key challenge in biology. This is especially difficult for variants found in the non-coding portion of the genome, which regulates when, where, and at which level genes are transcribed in each cell type. Gene regulatory instructions are encoded in units of 100bp- to 1kb-long DNA sequences called cis-regulatory elements (CREs). CREs such as enhancers and promoters contain binding sites for transcription factors (TFs), which function together with various transcriptional regulators and complexes to set the desired gene expression levels. This cis-regulatory code, the set of rules by which CRE sequences collectively control gene expression in a cell type, is incompletely understood, which makes it extremely challenging to predict how genetic variation alters gene regulation.

A comprehensive understanding of the cis-regulatory code would provide a blueprint of how cells differentiate into the various cell types during embryonic development, predict how genetic variants influence development and health, and identify the molecular mechanisms altered by disease-associated genetic differences. The resulting knowledge may also allow us to develop therapeutic interventions, including engineering enhancer variants with highly specific activities, to direct cells toward favorable gene expression programs that restore and maintain cellular function.

Deciphering the cis-regulatory code is an extraordinarily complex problem. Each cell type has a unique combination of TFs, expressed at specific levels and whose activity is sometimes controlled by extracellular signals. Given the activity levels of all TFs in a given cell type, the transcription of all genes should be predictable from the DNA sequence alone. However, such predictions are challenging since TFs can act combinatorially and influence multiple regulatory layers: TFs cooperate to bind and access CREs, change the chromatin environment around CREs, recruit other regulatory proteins, and, together with TFs at other CREs, regulate the transcription of target genes. Furthermore, many CREs can be read by specific TF combinations that may only be present in specific cell types. Thus, complexity arises from the many regulatory layers and the cell-type-specific nature by which TFs regulate these layers.

Although deciphering the cis-regulatory complexity is a daunting challenge, there has been tremendous progress (Figure 1). We have a much-improved understanding of gene regulation (Preissl *et al.*, 2023; Zeitlinger, 2020; Bonn and Furlong, 2008; Yáñez-Cuna *et al.*, 2013; Kim and Wysocka, 2023) and a vast and growing number of genomics data sets from multiple experimental assays over many cell types and conditions. Importantly, advances in computational methods, especially deep learning (Eraslan *et al.*, 2019), have shown that complex cis-regulatory rules can be learned from such data sets in a given cell type. These approaches increasingly reveal mechanistic insights into TF cooperativity, make experimental predictions, and allow the effect of genetic variants to be studied.

In this perspective, we delve into the current computational challenges facing the field of regulatory genomics, with a specific focus on the cis-regulatory code that regulates transcription (for post-transcriptional mechanisms, see (Keene, 2007; Zhao *et al.*, 2017)). We discuss state-of-the-art computational methods that aim to map CREs, learn the cis-regulatory rules of specific cell types, predict the effect of genetic variants, incorporate 3D chromatin organization to characterize long-range regulatory effects, and map gene regulatory networks during specific biological processes (Figure 1). We provide our perspective on the current gaps in knowledge, limitations of current methods and how to benchmark them, and opportunities for developing methods that analyze and integrate newly emerging data, such as those from spatial omics technologies. Finally, we outline possible strategies for closing the remaining gaps and creating a path toward a more general model of the cis-regulatory code that is more broadly applicable across cell types and individuals.

## Leveraging experimental progress: chromatin-based annotations of CREs

The advent of high-throughput genomic technologies has made it possible to comprehensively map the regulatory landscape across the genome in many cell types and conditions (Hawkins *et al.*, 2010; Zhou *et al.*, 2011). The basic building blocks of the cis-regulatory code, the TFs that bind to sequence motifs, can be mapped genome-wide inside cells using chromatin immunoprecipitation coupled to sequencing (ChIP-seq) or related assays. However, TF binding *in vivo* is highly cooperative and cell-type specific, and extensive ChIP experiments are only available for a few human cell types (Moore *et al.*, 2020). Therefore, although the binding specificity of the majority of human TFs has been experimentally determined (Lambert *et al.*, 2018; Rauluseviciute *et al.*, 2024), our understanding of the combinatorial landscape by which TFs cooperate to access and regulate different CREs across cell types is still limited.

To facilitate the discovery of CREs and their TF binding sites, a popular approach is to identify the genomic sequences that are accessible in chromatin. CREs have long been known to be hypersensitive to DNase digestion (Burch and Weintraub, 1983), and DNase-seq provides comprehensive and quantitative information on chromatin accessibility genome-wide (Song and Crawford, 2010; Thurman *et al.*, 2012). Another convenient assay to measure chromatin accessibility with less input material is ATAC-seq (Buenrostro *et al.*, 2015). However, while such assays allow the comprehensive identification of accessible regions in a cell type, which of these candidate CREs contribute to gene regulation under the examined conditions, and what class of CREs they might represent is unclear.

The classes of CREs can be broadly divided into those that tend to be constitutively open and those that are accessible in a cell-type-specific manner (Ernst and Kellis, 2013). Promoters, the regions around the transcription start sites that initiate transcription, as well as architectural elements that help organize chromatin in 3D (e.g., CTCF-bound regions), are two broad classes of CREs that tend to be constitutively open (Phillips and Corces, 2009). The cell-type-specific activation of promoters occurs through a class of CREs called enhancers, which can be located near the promoter or can contact the promoter from distal locations, e.g., up to multiple megabases away. Enhancers are open in a cell-type-specific fashion and thus can be identified through their differential accessibility. However, not all accessible CREs are active, and they may even have a repressive effect on the surrounding chromatin when harboring motifs for repressive TFs (Segert *et al.*, 2021; Pang and Snyder, 2020; Berest *et al.*, 2019). Other CREs may have a constitutive repressive effect, e.g., by more broadly promoting repressive chromatin marked by H3K27me3 or H3K9me3 (Thurman *et al.*, 2012; Delest *et al.*, 2012).

To better classify the function of candidate CREs, additional data sets that measure aspects of the chromatin state are helpful, including histone modifications measured by ChIP-seq. Specifically, active enhancers tend to be flanked by high histone acetylation levels and specific forms of histone methylation (e.g., H3K27ac, H3K4me1). Other markers for active enhancers are the presence of transcriptional co-activators (e.g., p300, Brd4, and Mediator) and enhancer transcription (e.g., measured by CAGE, PRO-seq, or NET-seq) (Policastro and Zentner, 2021). Obtaining comprehensive experimental data for characterizing chromatin states has been a major goal for the ENCODE (Moore *et al.*, 2020) and Roadmap Epigenomics (Roadmap Epigenomics Consortium *et al.*, 2015) projects and has opened the door to systematically analyzing CREs across cell types.

As data for multiple chromatin marks are often collected in the same cell type, a common strategy for systematically annotating candidate CREs is to integrate information using multivariate hidden Markov models such as ChromHMM, or related probabilistic models, to define chromatin states (Ernst and Kellis, 2010, 2012; Hoffman *et al.*, 2012; Libbrecht *et al.*, 2021). This strategy has been used for CRE annotations across a wide range of cell types and conditions (Roadmap Epigenomics Consortium et al., 2015). In addition to directly using observed data, CREs have been annotated in cell types with incompletely observed data using imputed epigenomic datasets (Ernst and Kellis, 2015; Schreiber *et al.*, 2020). The availability of data from hundreds of cell types and conditions has also spurred analysis approaches that can directly categorize different classes of cell-type restricted or constitutively active CREs (Meuleman *et al.*, 2020; Vu and Ernst, 2022).

While such chromatin-based approaches have increased the number of identified putative CREs into the millions (Meuleman *et al.*, 2020; Moore *et al.*, 2020), many additional CREs are likely encoded in the genome. For example, evolutionarily conserved sequence analyses suggest that a substantial portion of conserved non-coding bases are not well captured by large compendiums of annotations (Christmas *et al.*, 2023; Grujic *et al.*, 2020). Therefore, expanding these data sets by profiling rare or hard-to-access cell types and conditions is critical.

Single-cell chromatin assays are ideal for capturing CREs from rare cell types. Due to their sparsity and dimensionality, these datasets add additional computational challenges. The most widely applied single-cell chromatin assay is ATAC-seq, sometimes jointly performed with gene expression (Preissl *et al.*, 2023). Single-cell assays for measuring other epigenetic features, including histone modifications and DNA methylation, have also been developed and continue to mature, along with computational methods to integrate the resulting data (Preissl *et al.*, 2023; Shema *et al.*, 2019). To inform mechanistic models of gene regulation, another useful chromatin assay is single-molecule footprinting. This technology uses exogenous DNA methylation to capture the footprints of bound TFs and nucleosomes on the same DNA molecule, allowing absolute measurements of bound fractions and co-occurrence events in a population of cells (Kreibich *et al.*, 2023; Sönmezer *et al.*, 2021). In the future, assays that measure how chromatin accessibility, TF binding, and histone marks are spatially organized inside cells (Deng *et al.*, 2022; Lu *et al.*, 2022) could provide additional cell-type-specific information on gene regulation.

A continuing challenge is to benchmark CRE annotations. Traditionally, this is performed by comparing the results with established genome annotations and other experimental data not used during model learning (Ernst and Kellis, 2010; Vu and Ernst, 2022). However, this approach does not directly validate novel annotations. Additional high-throughput functional assays, such as massively parallel reporter assays (MPRA) and non-coding CRISPR-based screens, provide a promising avenue to evaluate and characterize CRE annotations (Gasperini *et al.*, 2020; Yao *et al.*, 2024).

## The solution to complexity: sequence-to-function models

The ultimate goal of understanding CRE function is to decipher the cis-regulatory code embedded within these sequences. Each assay measures specific regulatory activities across the genome in a given cell type, and these activities should be driven by TF binding motifs and possibly other sequence patterns within CRE sequences. However, determining the exact relationship between sequence and a given functional readout is difficult. The traditional approach is to select the regions with high TF binding, chromatin accessibility, or enhancer activity and to identify TF motifs that are statistically overrepresented (McLeay and Bailey, 2010). While this provides a set of TF motifs, it does not capture how the affinity and syntax of the motifs in their genomic context affect the readout (Crocker *et al.*, 2008; Farley *et al.*, 2016). Since each genomic region is very different, discovering genome-wide rules by which motifs interact and predicting the experimental readout is challenging. Fortunately, it is now possible to train neural networks ("sequence-to-function models") to perform this task.

During training, sequence-to-function models learn to predict an experimental readout across a large number of CREs directly from the underlying genomic sequence. By optimizing the prediction accuracy, the model learns sequence rules inside a 'black box' without prior biological assumptions. When the predictive performance is high on withheld data not seen by the model during training, this suggests that the learned rules apply genome-wide. To train on different experimental data types, models are typically optimized to the biological problem of interest. They may differ in their DNA input size, the selection of regions, model type (e.g., convolutional neural networks or transformers), architecture and model size (e.g., filters, layers, receptive field), and loss function. In general, training a model to predict high-resolution, high-coverage data quantitatively at base resolution produces the most nuanced sequence features (Avsec, Weilert, *et al.*, 2021; Toneyan *et al.*, 2022). For some purposes, models predict binary or categorical data or data that are averaged across genomic bins, which reduces the computing requirements. For example, by binning to 128 bp and using a transformer architecture, Enformer predicts data across 200 kb (Avsec, Agarwal, *et al.*, 2021; K. M. Chen *et al.*, 2022; Kelley, 2020; Kelley *et al.*, 2018).

The high prediction accuracy of these models is useful on their own, e.g., to test the effect of genetic variants, but considerable power of sequence-to-function models lies in their interpretation. This is counterintuitive as 'black box' models are traditionally considered as uninterpretable because features and relationships are learned in a distributed manner inside the neural network. However, with DNA sequence being a relatively simple input, interpretation approaches that query the model as a whole have been very successful in revealing sequence motifs and their syntax rules (Novakovsky *et al.*, 2023; Alipanahi *et al.*, 2015; de Almeida *et al.*, 2022; Avsec, Weilert, *et al.*, 2021). This suggests that DNA 'black box' models are interpretable, at least to the extent that the rules follow somewhat expected patterns. An alternative is to incorporate *a priori* biological knowledge into the neural network architecture, but these constraints come at the expense of learning more complex phenomena and cannot capture unknown cis-regulatory sequence rules (e.g.,(Agarwal *et al.*, 2021; Novakovsky *et al.*, 2023; Balcı *et al.*, 2023). Therefore, interpreting 'black box' models is currently the best approach to uncover new sequence rules in the cell type of interest.

There are several complementary approaches by which models can be interpreted. The most common first step is to use an attribution method (e.g., DeepLIFT (Shrikumar *et al.*, 2019) or Deep SHAP (Lundberg and Lee, 2017)), which assigns scores for how much each feature in the input (i.e., each base) contributes to the output prediction. Motifs highlighted in genomic regions by high scores can be summarized by tools like TF-MoDisco (Shrikumar *et al.*, 2020). These TF motif representations can then label high-scoring motif instances in the genome. This approach outperforms traditional position weight matrix (PWM) methods (Avsec, Weilert, *et al.*, 2021) because motif instances mapped in this way were informative to the model in the specific genomic sequence context.

To extract additional sequence rules, such as how TF motifs interact, trained models can be queried with *in silico* sequence designs, e.g., by perturbing motifs in genomic sequences or injecting motifs into randomized sequences (Zhou and Troyanskaya, 2015; Alipanahi *et al.*, 2015; Avsec, Weilert, *et al.*, 2021; Trevino *et al.*, 2021; Nair *et al.*, 2022; de Almeida *et al.*, 2022; Koo *et al.*, 2021). Analyzing predictions with systematic sequence designs allows the extraction of relative motif affinities (Alexandari *et al.*, 2023; Brennan *et al.*, 2023) and specific syntax rules by which motif pairs interact (Avsec, Weilert, *et al.*, 2021; de Almeida *et al.*, 2022; Koo *et al.*, 2021).

In this manner, the genome-wide cis-regulatory sequence rules that underlie various data modalities have been characterized and linked to molecular mechanisms (Novakovsky *et al.*, 2023). For example, the syntax rules by which TFs cooperate based on *in vivo* binding data show that some TFs preferentially interact when the motifs are within nucleosome distance, while others may physically cooperate on DNA when the motifs are spaced at a fixed distance (Avsec, Weilert, *et al.*, 2021). These rules match prior mechanistic studies (Smith *et al.*, 2023; Long *et al.*, 2016), showing that neural networks can learn accurate biological representations without *a priori* knowledge of the underlying biophysical properties and mechanistic principles.

Sequence-to-function models can also predict or interpret how TF binding relates to cell-type-specific chromatin environments. Some approaches incorporate chromatin features alongside sequence as model inputs, allowing the models to distinguish between direct sequence predictors of TF-DNA binding and a more generalized dependency on chromatin state (Srivastava *et al.*, 2021; Arora *et al.*, 2023). With a sufficiently complex sequence-to-function model, chromatin accessibility and other chromatin features can themselves be predicted from DNA, revealing the sequence rules by which TFs shape the chromatin landscape. Such approaches have revealed that TFs drive chromatin accessibility proportional to the motif affinity, that some TFs have a repressive effect, and that TFs often function synergistically in making chromatin accessible (D. S. Kim *et al.*, 2021; Brennan *et al.*, 2023; Bravo González-Blas *et al.*, 2024). Presumably, these sequence rules reflect how TFs act on nucleosomes (Brennan *et al.*, 2023), but the mechanisms for this type of TF cooperativity are not well understood. Here is therefore an opportunity for sequence models to inspire mechanistic studies.

Another challenge is understanding how sequence rules specify enhancer activity and target gene activation. The rules of enhancer activity differ from those of chromatin accessibility (Brennan *et al.*, 2023), but how exactly is not well understood. Models trained on large-scale reporter assays have shown that motif syntax and repressive motifs are important for enhancer activity (de Almeida *et al.*, 2022; Movva *et al.*, 2019). These assays are however typically episomal and use short DNA regions (Inoue *et al.*, 2017), thus it would be beneficial to obtain additional insights into enhancer activity. For example, ChIP-seq signals of histone modifications typically flank active enhancers in the genome and can be predicted from DNA sequence (Avsec, Agarwal, *et al.*, 2021; K. M. Chen *et al.*, 2022; Kelley, 2020; Kelley *et al.*, 2018), but no careful model interpretation has been performed to understand the underlying cis-regulatory sequence rules.

Ultimately, one would like to directly predict gene expression, either steady-state RNA levels or nascent transcription data, from DNA sequence. This is possible and yields highly accurate predictions (Avsec, Agarwal, *et al.*, 2021; Kelley *et al.*, 2018; Linder *et al.*, 2023) and promoter syntax (Dudnyk *et al.*, 2024; He and Danko, 2024; Cochran *et al.*, 2024). However, the input from distal enhancers is not well captured, suggesting that the models are missing some cell-type-specific sequence rules, perhaps related to enhancer activation or long-distance enhancer-promoter interactions (Karollus *et al.*, 2023; Kathail *et al.*, 2024).

This shows that our understanding of the cis-regulatory code and the molecular mechanisms by which TFs mediate enhancer activation and target gene expression is still incomplete. Current models specifically learn the data on which they were trained and are thus specific for a data modality and cell type. There is no straightforward way to combine sequence rules from different models coherently. One solution may be to train many data modalities as a multi-task model (Avsec, Agarwal, *et al.*, 2021; K. M. Chen *et al.*, 2022; Kelley, 2020; Kelley *et al.*, 2018), but this does not necessarily mean that the learned sequence rules are more coherent and better represent biology.

A solution for handling different data modalities is to learn assay biases in separate deep learning models (e.g., Tn5 insertion bias in ATAC-seq data) such that specific features of the cis-regulatory code can be learned more explicitly (Brennan *et al.*, 2023; Pampari *et al.*, 2023). If the biophysical properties of these rules are known, secondary surrogate models can be trained to fit these properties, e.g. TF binding and cooperativity (Seitz *et al.*, 2024). This could eventually lead to biophysical models of the cis-regulatory code, but this would require extensive knowledge of the molecular mechanisms beyond transcription factor binding, which currently does not exist.

To improve existing models or develop new models, a better mechanistic understanding of the cis-regulatory code through systematic interpretation of various models would be highly beneficial. While interpreting models to uncover new biology, it may also be important to examine the model's limitations, e.g., using data simulation models (Chen and Capra, 2020; V. Chen *et al.*, 2022; Prakash *et al.*, 2021), and to analyze the limitations of the experimental data, e.g., assay biases, experimental artifacts, or the effect of low resolution or low coverage. If this is done for many data sets and modalities, the sequence rules should overlap and provide clearer expectations of the underlying biophysical constraints and molecular mechanisms that lead to the activation of enhancers and their target genes. The ultimate challenge will be to train models that generalize cis-regulatory rules and can predict data for cell types not trained on, a goal that may require significant computational innovation involving domain adaptation.

Meanwhile, experimental validation is needed to ensure the learned sequence rules are accurate. One approach is to predict and test the effect of targeted perturbations, e.g., mutating genomic regions by CRISPR (Avsec, Weilert, *et al.*, 2021) or knocking down a TF (Brennan *et al.*, 2023). Large-scale MPRA reporter assays can be used to validate the learned sequence rules at higher throughput (D. S. Kim *et al.*, 2021). A powerful validation is to use the trained model to create synthetic enhancers and test them *in vivo* with a reporter assay. Synthetic designs can be generated from random sequences, manual manipulation or trimming of existing enhancers, or through *de novo* design of enhancers (de Almeida *et al.*, 2024; Taskiran *et al.*, 2024). In the future, such synthetic enhancer designs could be used to create enhancers with increased or altered function, with the potential of using such designs for therapeutic treatments.

# From genotype to phenotype: predicting the effect of regulatory variants

A promising application of sequence-to-function models is the prediction and interpretation of regulatory variants involved in the predisposition, onset or progression of complex diseases. Most SNPs identified in genome-wide association studies (GWAS) fall in the non-coding portion of the human genome (Buniello *et al.*, 2019). However, due to linkage disequilibrium, the SNPs identified in GWAS are often not the causal variants but are located nearby. This necessitates fine-mapping to pinpoint the causal variants, most of which are expected to alter gene expression. Sequence-to-function models can be used to identify and interpret regulatory variants by quantifying their predicted effect on expression variation or other molecular features.

Different types of gene expression variation are however not equally amenable to modeling. Predicting gene-to-gene variation within the same cell type is an easier task because the levels vary widely and can be predicted from promoter-proximal sequences without a comprehensive understanding of the cis-regulatory code of distal enhancers (Karollus *et al.*, 2023). Predicting the variation between cell types across an organism is more challenging because the cis-regulatory code is highly diverse across cell types and often driven by distal enhancers far away from promoters. Predicting variation in gene expression across individuals in a population is the most challenging task (Sasse *et al.*, 2023; Huang *et al.*, 2023; Tang *et al.*, 2023). Not only does it require predicting cell-type-specific gene expression of individual genetic variants, whose effects are often small, but also how they affect specific gene expression programs and phenotypes of cells, leading to disease susceptibility (Manolio *et al.*, 2009). Only very few examples exist where this is well characterized (e.g., (Claussnitzer *et al.*, 2015)).

A starting point for this challenging task is to assess which genetic variants differentially affect TF binding. The simplest models use PWMs to assess how TF motif affinities differ between the alternate and the reference alleles (Kumar *et al.*, 2017; Fornes *et al.*, 2018; Santana-Garcia *et al.*, 2019). Predictions from such methods can correlate well with observed allele-specific binding events derived from ChIP-seq experiments (Fornes *et al.*, 2018). More sophisticated models of TF binding specificity are trained on high-throughput data, such as *in vitro* TF binding data (Martin *et al.*, 2019). Another category of methods combines multiple features, such as evolutionary conservation, chromatin states, and enhancer-promoter interactions, to predict causal variants. Such models are trained or evaluated on known genetic variants from the Human Gene Mutation Database or ClinVar (Gao *et al.*, 2018; Huang *et al.*, 2017; Rogers *et al.*, 2018) and can be combined into ensemble models (Zhang *et al.*, 2019).

If sequence-to-function models are trained on molecular genomic data sets, they can directly predict how genetic variants alter the experimental outcome (Sokolova *et al.*, 2024). This is because these models learn genome-wide rules and make accurate predictions across diverse genomic sequences, including variants not directly trained on. Models that predict TF binding, chromatin accessibility, and histone modifications are very accurate in this regard (Zhou and Troyanskaya, 2015; Kelley *et al.*, 2016; Trevino *et al.*, 2021; Z. Chen *et al.*, 2023). However, it is not always clear how these activities drive enhancer activation and specific changes in nearby promoter activity. Therefore, sequence-to-function models that predict MPRA reporter activity (Movva *et al.*, 2019) and expression data are appealing (Zhou *et al.*, 2018; Avsec, Agarwal, *et al.*, 2021; Linder *et al.*, 2023). While these models perform well on promoter and splicing variants, their performance is limited when predicting cell-type-specific effects involving long-range enhancer-promoter interactions (Karollus *et al.*, 2023). Thus, current models still show gaps in predicting the effect of genetic variants on gene expression. However, due to their ability to predict which mutations have an effect, they are increasingly used to identify and interpret regulatory variants.

For studying complex human disease, the additional information gained from sequence-to-function models have so far been limited based on multiple evaluations in the context of GWAS data (Dey *et al.*, 2020). This may change with more advanced sequence-to-function models, the availability of more data, and better strategies for integrating the model predictions with human GWAS data. A major obstacle is that the model predictions are specific for the cell type of the training data, and thus in many cases do not cover the cell types relevant for the GWAS traits. A potential bridge between genetic variants and disease traits is gene expression data or other genomics data profiled across individuals, ideally from a tissue of interest for the trait (Drusinsky *et al.*, 2024). These data can be used to infer molecular quantitative trait loci (QTLs) (e.g.,(Ramdas *et al.*, 2022)). While there has been limited overlap between expression QTLs and GWAS hits (Mostafavi *et al.*, 2023), additional molecular QTLs have shown greater though still partial overlap (Wu *et al.*, 2023). Overall, identifying and interpreting causal genetic variants from GWAS studies is still a major challenge.

A potentially lower-hanging fruit is to use sequence-to-function models to identify genetic variants that cause rare diseases (Sokolova *et al.*, 2024). Rare variants are likely purged from the population by purifying selection and thus can have larger effect sizes that are more easily detected. Since they are rare in the population, GWAS studies lack the power to discover them, but they can still be predicted by sequence-to-function models.

Moving forward, a major bottleneck is the availability of uniformly processed and validated genomics data, as well as high-quality QTL data for expression and other genomics data sets for less characterized cell types. But even for well-studied cell types, identifying the key genes that affect the disease phenotype in the presence of multiple regulatory variants and secondary transcription effects is challenging (Manolio *et al.*, 2009; Liu *et al.*, 2019; Li and Ritchie, 2021). Models that predict the effect of multiple regulatory variants and consider coding and non-coding epistasis in predicting disease outcomes would be useful (Monti and Ohler, 2023). We note that while models are generally able to make accurate predictions for unseen variants, it is nevertheless important to include more ancestry-diverse sequences during training and to benchmark variants from the entire population so that all humans can maximally benefit in eventual clinical applications (Martin *et al.*, 2017; Taylor *et al.*, 2024). Finally, it will be critical to further develop and apply more experimental techniques such as CRISPR editing to validate causal genetic variants (Z. Chen *et al.*, 2023; Pihlajamaa *et al.*, 2023).

# Filling the gap: 3D genome organization and long-range regulatory interactions

One of the outstanding problems faced by gene expression models is to capture and integrate the effect of multiple CREs, proximal and distal, on each gene. This is challenging because enhancers can act over large genomic distances of over 1Mb to modulate the expression levels of their target genes (Chandra *et al.*, 2021; Long *et al.*, 2020; Lettice *et al.*, 2003). Since these regulatory connections are thought to require physical proximity in 3D space, mapping the 3D organization of DNA in the context of chromatin could detect important distal interactions and improve our ability to predict gene expression from DNA sequence.

The 3D chromatin structure can be captured as genome-wide contact maps using Chromosome Conformation Capture (3C) technologies such as Hi-C (Lieberman-Aiden *et al.*, 2009; Rao *et al.*, 2014). These methods can characterize multiple layers of genome organization, including cell-type-specific aspects (Dekker *et al.*, 2023). However, the limited resolution of Hi-C (typically 5kb-40kb) and the requirement for very high sequencing coverage (e.g. 1 billion for 5 kb resolution) coupled with underrepresentation of distal interactions make it challenging to assign enhancers to target genes. This led to the development of additional steps in the assay that increase the coverage of the relevant regions (Mumbach *et al.*, 2016; Fang *et al.*, 2016; Mifsud *et al.*, 2015; Fullwood *et al.*, 2009) or increase the resolution by which contacts are detected. For example, Micro-C can achieve kilobase resolution genome-wide (Hsieh *et al.*, 2015; Krietenstein *et al.*, 2020; Harris *et al.*, 2023) or sub-kilobase resolution for targeted regions (Goel *et al.*, 2023). High-resolution contact maps in primary cell types have helped predict target genes for distally located disease-associated genetic variants (Chandra *et al.*, 2021; Hamley *et al.*, 2023; Javierre *et al.*, 2016). However, they are difficult to obtain for a large collection of primary cell types, and the genome-wide resolution is still not at the level of individual CREs to obtain generalizable insights into enhancer-promoter interactions.

Computational methods that detect patterns in these contact maps have revealed multiple levels of organization that could influence gene expression: chromatin compartments, topologically associating domains (TADs), and chromatin interactions or loops (Zhang *et al.*, 2024). Multi-megabase chromatin compartments correspond to the large-scale division between transcriptionally active euchromatin (A compartment) and inactive heterochromatin (B compartment). TADs spanning 100kb-1Mb regions are often considered as regulatory units of coordinated gene expression within which CREs interact more frequently with one another (Beagan and Phillips-Cremins, 2020). Two convergent CTCF motifs, which stop cohesin-mediated loop extrusion (Fudenberg *et al.*, 2016), often demarcate their boundaries. Another pattern are preferential interactions among CREs, whether it is mediated by loop extrusion to specifically bring two distal CREs in close proximity (e.g., chromatin loop between an enhancer and a promoter) or broader colocalization of CREs in the 3D space potentially through their homotypic interactions.

With the increasing resolution of contact maps, recent methods have detected additional patterns that provide insights into how CREs influence loop extrusion and, in turn, gene expression (Vian *et al.*, 2018; Yoon *et al.*, 2022; Guo *et al.*, 2022). For instance, stripes form when a loop anchor interacts with a long stretch of chromatin at high frequency (Yoon *et al.*, 2022). Stripe anchors generally mark clusters of enhancers that regulate multiple genes throughout the domain (Vian *et al.*, 2018), e.g. at immunoglobulin loci (Hu *et al.*, 2023; Vian *et al.*, 2018) or developmentally regulated genes (Kraft *et al.*, 2019). However, how these emerging contact patterns can be generalized to model gene regulation is not yet clear.

To understand how 3D genome organization is instructed by DNA sequence, sequence-to-function models have been trained to predict Hi-C contact maps (Zhang *et al.*, 2024; Zhou, 2022; Fudenberg *et al.*, 2020; Schwessinger *et al.*, 2020). Interpretation of these models suggests that the backbone of 3D organization is established by motifs of architectural proteins such as CTCF and tends to be cell-type invariant (Piecyk *et al.*, 2022). So far, these models have not provided novel sequence features that promote specific enhancer-promoter interactions. It is possible that further improvements, e.g., using chromatin accessibility and ChIP-seq data as additional input during training (Tan *et al.*, 2023), better data coverage and resolution, or more extensive model interpretation, could reveal cell-type-specific features that help predict gene expression.

The role of chromatin contacts in determining the effect of an enhancer on gene expression has also been analyzed more explicitly (Fulco *et al.*, 2019). This led to the Activity-by-Contact (ABC) model, which has been applied across a large collection of cell types to better link non-coding risk variants to disease genes (Nasser *et al.*, 2021). The ABC model assumes that the influence of an enhancer depends on its activity multiplied by the intensity of contact with the promoter and that multiple enhancers contribute to gene expression in an additive manner. This analysis demonstrated, in part, the utility of cell-type-specific contact maps in determining functional enhancer-promoter interactions but also showed that genomic distance is the strongest determinant of enhancer-promoter contacts and has a large effect on gene expression levels. This distance dependence agrees with recent experimental data (Zuin *et al.*, 2022).

While the ABC model serves as a good baseline model, it cannot predict the effect of some validated enhancers, suggesting that additional unknown mechanisms are at play. For example, closely spaced enhancers may function synergistically or redundantly, and specific promoters may not be as responsive to enhancers (Liu *et al.*, 2019; Gschwind *et al.*, 2023). Most notably, it is unclear how some enhancers can find their target genes with high specificity over hundreds of kilobases, while others cannot. Studies in the fruit fly suggest a new class of CREs that enables enhancers to do so by mediating long-distance chromatin interactions (Batut *et al.*, 2022). Such "extender elements" have recently been identified in mice: when located next to an enhancer, they allow the enhancer to regulate target genes over hundreds of kilobases of distance (Bower *et al.*, 2024).

Taken together, these results suggest that enhancer-promoter interactions are a complex layer of the cis-regulatory code. Enhancer-promoter interactions depend on genomic distance, the 3D organization created by architectural CREs, and long-range contacts enabled by newly emerging CREs. Predictive features might be identified and characterized more precisely by additional experimental efforts, including generating high-resolution contact maps for more cell types, devising experimental techniques with improved temporal and spatial resolution ("4D" (Sekelja *et al.*, 2016; Dekker *et al.*, 2023), measuring multiple modalities such as chromatin organization and gene expression simultaneously (Zhou *et al.*, 2024; Liu *et al.*, 2021, 2023; Su *et al.*, 2020), as well as mapping multi-way contact of chromatin (Quinodoz *et al.*, 2018; Beagrie *et al.*, 2017; A. S. Deshpande *et al.*, 2022; Tavares-Cadete *et al.*, 2020; Oudelaar *et al.*, 2022; L.-F. Chen *et al.*, 2023). Significant computational innovation will be required to leverage these additional data to extract sequence features that are currently missing or create models that successfully learn long-distance interactions and the interplay of multiple CREs in gene regulation.

## Assembling the parts: gene regulatory networks

Ultimately, cis-regulatory sequences are not only key for predicting expression levels, but also how cells change their gene expression program dynamically during embryonic development, exposure to stress, or disease pathogenesis. The methods described so far aimed to predict gene expression given a fixed steady-state cellular state defined by a specific set of active TFs. To predict how cells change their gene expression program, we need to understand how TF activities change over time. The activity of some TFs is regulated by signal transduction pathways in response to extracellular stimuli. However, most TFs, and the regulators they depend upon, are, to some extent, regulated at the expression level. Thus, TF activities are themselves the target of CRE regulation, which creates a dynamic system that allows cells to transition along specific cellular trajectories depending on their extracellular environment. The changing interactions between TFs, CRE activity, and the expression of target genes over time are called gene regulatory networks (GRNs).

GRNs play an important role during embryonic development, where cells transition through multiple states to eventually acquire a specific cell fate with a characteristic cell-type-specific expression program. Indeed, the concept of GRNs was pioneered in sea urchin and Drosophila embryos by studying how key regulators identified through developmental genetics are themselves regulated (Levine and Davidson, 2005). This led to the discovery of enhancers, which each drive the expression of the target gene in a specific spatio-temporal manner and are controlled by the combinatorial input from TFs active at that time. By tracing back how these TFs are regulated, coherent descriptions of developmental processes were obtained. However, such top-down models were restricted to key enhancers and TFs, required a laborious iterative experimental process, and the identified cis-regulatory sequence rules did not generalize to allow genome-wide predictions of gene expression from sequence alone.

Methods aimed at building GRNs from genome-wide data by explicitly modeling sequence motifs and their TFs as regulators are called cis-GRN methods. These methods model gene expression over time or across cell types as a function of TF motifs found in candidate CRE regions, identified by histone marks, chromatin accessibility, or TF ChIP-seq data taken from developmental model systems (González *et al.*, 2015; Ding, Hagood, *et al.*, 2018; Siahpirani *et al.*, 2022). The initial candidate TF motifs are typically identified by scanning CREs with a library of known motifs (Sherwood *et al.*, 2014; Chen *et al.*, 2017; Bentsen *et al.*, 2020) or by *de novo* motif finding (Setty and Leslie, 2015). Relevant TF motif features are then discovered through their association with co-regulated target genes or modules. A drawback of this modeling approach is that it strongly depends on the choice of candidate CRE regions, TF motifs, and target gene modules. Initial methods have focused on proximal enhancers near promoters, but recent approaches also model or incorporate distal enhancers, either by their high accessibility when the target is active (e.g. (Bravo González-Blas *et al.*, 2023)) or by using 3C data (e.g., (Karbalayghareh *et al.*, 2022)). An advantage of cis-GRN methods is that CREs and the TF motifs involved in a particular process can be identified *de novo* and followed up with experiments.

An alternative set of approaches for building GRNs are trans-GRN methods, which infer the role of TFs and other "trans" regulators through their co-expression with their target genes (Amit *et al.*, 2011; Kim *et al.*, 2009). These methods require large sample sizes and use probabilistic graphical models such as Bayesian networks and their extensions (Friedman *et al.*, 2000; Segal *et al.*, 2003) or dependency networks with linear (Greenfield *et al.*, 2013; Siahpirani and Roy, 2017) or non-linear regression models (Huynh-Thu *et al.*, 2010; Baran *et al.*, 2012). Several of these methods have integrated known TF motifs located near promoters as secondary features to inform the network structure (Greenfield *et al.*, 2013; Siahpirani and Roy, 2017; Petralia *et al.*, 2015; Glass *et al.*, 2013). TF motifs can also inform TF activity levels when the expression levels are a poor predictor of its activity, e.g., because the TF is regulated by signaling (Miraldi *et al.*, 2019; Wang *et al.*, 2018).

The rise in single-cell genomics technology has greatly benefitted GRN inference, especially trans-GRNs, which can leverage the large amounts of single-cell expression experiments (i.e., scRNA-seq) from normal and disease conditions. Single-cell resolution data better distinguish cell types and states among heterogeneous biological samples, thereby improving the discovery of the TFs that define each cell type. The larger sample sizes also allow for capturing additional non-linearities between TF and target expression with the help of deep learning models (Shu *et al.*, 2021; Luo *et al.*, 2022). Furthermore, cellular dynamics can be inferred using measured time (e.g. (Ding, Aronow, *et al.*, 2018)), velocity (Bocci *et al.*, 2022; Burdziak *et al.*, 2023), or pseudotime (L. Wang *et al.*, 2023; J. Kim *et al.*, 2021; A. Deshpande *et al.*, 2022), which can be used to inform GRN inference algorithms to capture fine-grained dynamics. These inferred networks can be analyzed to identify regulators that may mediate the transition between specific cell states, e.g., by determining rewiring scores of its local network topology (e.g., (Zhang *et al.*, 2023; L. Wang *et al.*, 2023) or using *in silico* perturbation analysis (Kamimoto *et al.*, 2023; Fleck *et al.*, 2023). These predictions can be used to engineer improved *in vitro* differentiation or trans-differentiation models.

Additional power for both cis and trans-GRNs comes from the combination of scRNA-seq with single-cell ATAC-seq (scATAC-seq) data, ideally from a multi-omics assay where these data modalities are measured simultaneously in the same cells (Badia-I-Mompel *et al.*, 2023). Single-cell chromatin accessibility data improve the quality of the inferred networks, resolution of cell types and allows the analysis of TF motifs within CREs (L. Wang *et al.*, 2023; Zhang *et al.*, 2023; Bravo González-Blas *et al.*, 2023). Especially for cis-GRNs, cell-type-specific accessibility changes enable the prediction of long-range enhancer-promoter interactions without requiring Hi-C experiments (Pliner *et al.*, 2018; Sakaue *et al.*, 2024; Mitra *et al.*, 2024).

Through such recent advances, trans-GRN and cis-GRN approaches are increasingly being combined, but current approaches still primarily leverage one approach and are limited in the information they capture. For trans-GRN methods, CREs and their sequence information are a secondary feature, which limits their ability to predict the effect of genetic variation and reveal the molecular underpinnings of the cis-regulatory code. But without the "trans" information, cis-GRNs cannot directly infer which TFs and additional cell-type-specific regulators shape the TF activities that regulate the identified TF motifs. Furthermore, both approaches typically lack the TF motif interaction rules and predictive accuracy that sequence-to-function models provide. Combining different approaches in a more seamless way could therefore enable an improved understanding of the cis-regulatory code.

A hurdle towards this goal is the benchmarking of cell-type-specific GRN models beyond well-studied systems. For example, comprehensive gold standards are lacking when studying human cell types not covered by ENCODE (Pratapa *et al.*, 2020; McCalla *et al.*, 2023; Chen and Mar, 2018). To confirm the identity of a regulator and establish a relationship with target genes as causal therefore requires experimental perturbations (Seçilmiş *et al.*, 2022; Choo *et al.*, 2024). Since such follow-up experiments can be time-consuming, the availability of high-throughput perturbation screens that measure scRNA-seq in response to various regulator perturbations in individual cells (Dixit *et al.*, 2016; Replogle *et al.*, 2022; Schraivogel *et al.*, 2023) would accelerate this validation step. Such data could also improve the models, including their ability to predict how perturbations disrupt cellular function. Finally, centralized repositories of datasets and methods (Ben Guebila *et al.*, 2023; Wen *et al.*, 2023; Chevalley *et al.*, 2023) will be useful to advance methodological development and assess their practical utility for inferring the cis-regulatory code.

# Multi-scale integration: spatial transcriptomics and beyond

As we improve our ability to integrate different data modalities into coherent computational frameworks, an exciting new frontier will be the integration of data across multiple scales, from molecule to cell to tissue and organs. A particularly interesting aspect of such multi-scale integration is the spatial organization of cells, which determines which signals a cell receives from its neighboring cells, an element of GRNs that can currently only be inferred indirectly. Spatial aspects of gene regulation have long been studied at a small scale during embryonic development (Dubuis *et al.*, 2013) and are particularly relevant in the brain (Piwecka *et al.*, 2023). Such spatial organization can now be captured in a high-throughput manner for virtually any tissue using recently developed spatial transcriptomics technologies.

Spatial transcriptomics quantifies the expression distributions of large numbers of genes in a tissue at the resolution of single (or few) cells or individual transcripts (Moses and Pachter, 2022). These technologies rely either on sequencing with spatial barcoding (A. Chen *et al.*, 2022; Rodriques *et al.*, 2019) or single-molecule-fluorescent in situ hybridization (Chen *et al.*, 2015; Lubeck *et al.*, 2014). All methods aim to provide a comprehensive description of gene expression patterns across cells and tissues, without requiring prior hypotheses on which genes might be differentially expressed. However, current methods have limitations and represent different tradeoffs between the number of genes measured, spatial resolution, and tissue area coverage (Moses and Pachter, 2022).

While these methods are still in their infancy, they have created new opportunities for developing analytical tools that can extract different types of biological patterns. Such computational tools are capable of (1) identifying genes that exhibit interesting spatial patterns (Svensson *et al.*, 2018), (2) distilling a large number of spatial expression patterns into a smaller, representative set of patterns (Townes and Engelhardt, 2023), (3) inferring prominent spatial regions in the tissue (Hu *et al.*, 2021; Dong and Zhang, 2022), (4) characterizing cell-cell interactions in terms of involved cell types or genes (Arnol *et al.*, 2019; Dries *et al.*, 2021; Cang *et al.*, 2023), (5) identifying gene-gene interactions, e.g., ligand-receptor pairs, related to spatial expression (Yuan and Bar-Joseph, 2020; Tanevski *et al.*, 2022), and (6) detecting transcript localization (Xia *et al.*, 2019; Mah *et al.*, 2024) or transcript co-localization (Kumar *et al.*, 2024) at subcellular resolution. With these promising developments, it will be important to systematically evaluate and benchmark these tools (Moses and Pachter, 2022).

Spatial transcriptomics promises to provide novel insights into how the cellular dynamics and organization of tissues influence chromatin organization and gene regulation, but their potential has so far remained largely untapped. Glimpses into what is possible can be seen in some existing approaches, e.g., those for detecting cell-cell communication while incorporating signaling and regulatory networks (Browaeys *et al.*, 2020) or identifying tissue-level variations in RNA localization events that hint at post-transcriptional regulatory processes (Kumar *et al.*, 2024).

In the future, integrating spatial transcriptomics with additional data could improve the identification of signaling events and their impact on gene regulation in specific tissues and cell types. For example, combining spatial transcriptomics with spatial chromatin accessibility assays (Lu *et al.*, 2022) could help understand gene regulatory mechanisms. Furthermore, tools for mapping non-spatial single-cell data to spatial data from the same tissue (Biancalani *et al.*, 2021) could be highly informative, as they can lead to a common analytical framework for analyzing different single-cell measurements, e.g., transcripts and chromatin accessibility (Fang *et al.*, 2021), contact maps (Rappoport *et al.*, 2023), and proteomics (Bennett *et al.*, 2023).

# Future outlook for deciphering the cis-regulatory code

While there is rapid progress, major bottlenecks still exist in four areas. (1) There are clear gaps in our mechanistic understanding of the cis-regulatory code (Figure 1, blue). (2) We need experimental data for more cell types and more comprehensive multiomic data sets, including perturbation experiments, to better model all steps of the cis-regulatory code (Figure 1, yellow). (3) We need better GRN methods that more seamlessly combine cis-GRN, trans-GRN, and sequence-to-function approaches to model cell state changes (Figure 1, green). (4) We need major advances in sequence-to-function models that leverage the innovations above into more integrated and generalizable frameworks (Figure 1, pink).

Since we have an incomplete mechanistic understanding of how the cis-regulatory code is executed from sequence all the way to gene expression, it is currently difficult to pinpoint which regulatory steps are not well captured by current sequence-to-function expression models. Most notably, it is unclear how multiple enhancers activate specific target genes, which could depend on their biochemical activities, chromatin environment, relative distances, 3D organization, and other nearby CREs. Approaches that integrate 3C data, single-cell multiomic data, single-molecule footprinting, and perturbation experiments could provide the much-needed insights. The steps before, the local activities produced by TFs at enhancers, leading to chromatin accessibility, histone modifications, and other activating or repressing biochemical properties, are also not well understood. Deciphering the steps and general mechanistic principles will require an iterative process of model interpretation and experimental testing. Altogether, the mechanistic insights will help benchmark current models and enable more focused efforts to improve them.

Another challenge will be to predict gene expression across a much larger number of cell types. Each cell type has a unique combination and TF activities, and the exact sequence rules by which the TFs read out CREs cannot easily be predicted based on the TFs' individual binding specificities (Jolma *et al.*, 2015) and will require more high-resolution *in vivo* TF binding data, perhaps by adopting large-scale approaches (Perez *et al.*, 2023). Increasing the experimental coverage to hard-to-access cell types and conditions during developmental processes and in heterogeneous adult tissues will likely occur through more general methods such as single-cell multiomics data. While some missing data can be imputed when appropriate training data exist (Ernst and Kellis, 2015; Schreiber *et al.*, 2020), very unique combinatorial TF binding specificities are difficult to discover without sufficient experimental data.

One way to obtain missing cis-regulatory sequence information without experimental data is to directly leverage the large number of sequenced genomes across species. The specific combinations of TFs that specify a cell type tend to be evolutionarily conserved (Tarashansky *et al.*, 2021; Kuderna *et al.*, 2024), allowing CREs to be studied across evolution with sequence-to-function models (Kaplow *et al.*, 2023; Minnoye *et al.*, 2020). Furthermore, DNA large language models trained to predict masked genome sequences may detect combinatorial TF motif patterns in some cases (Karollus *et al.*, 2024; Silva *et al.*, 2024) but currently struggle to learn this information when trained on genome-wide mammalian sequence data (Tang and Koo, 2024). Future models trained more specifically on cis-regulatory regions could capture combinatorial TF binding specificities and thus complement information learned from sequence-to-function models.

Another important gap is GRN methods that more fully describe how cells dynamically change their TF repertoire and gene expression program. The increasing number and quality of single-cell spatial and temporal multiomics data create an opportunity to develop methods that better integrate cis-GRN, trans-GRN and sequence-to-function approaches. This includes trans-GRN methods that link expression programs to the cell-type-specific distal and proximal enhancers that regulate the genes (Mitra *et al.*, 2024; Sakaue *et al.*, 2024), sequence-to-function models that identify the effect of TF motifs (D. S. Kim *et al.*, 2021; Bravo González-Blas *et al.*, 2024), and cis-GRN methods that more explicitly identify which TF might bind these motifs based on expression data (Yang and Pe'er, 2024). Such methods can describe cellular changes across time and tissues and point to key TFs and CREs (Maslova *et al.*, 2020; Janssens *et al.*, 2022; Özel *et al.*, 2022).

As with sequence-to-function models, the insights obtained from modeling GRNs are however specific to the system of interest. Although the principles should conceptually extend to other biological systems, there is currently no generalizable framework that enables the effective prediction of expression changes *de novo*, e.g., based on only TF activities or accessible CREs. Such domain adaptation will require significant computational innovation. Intermediate steps towards this goal may involve developing sequence-to-function methods that learn more directly how enhancers change their activities as a function of changing TF activities.

In the long term, fully deciphering the cis-regulatory code will be important to predict how genetic variation affects multi-scale behavior at the organismal, cellular and molecular level. We will need major innovations in computational models to accurately predict cell-type-specific expression variation and the effect of genetic variants. Knowledge about the mechanistic steps and structural constraints of the cis-regulatory code could be introduced, for example, through geometric deep learning (H. Wang *et al.*, 2023; AlQuraishi and Sorger, 2021; Bronstein *et al.*, 2021), while foundation models that can learn from large amounts of non-coding genome sequences could improve generalization to new cell types and systems (Szałata *et al.*, 2024; Simon *et al.*, 2024). Finally, it is expected that more experimental data, including high-throughput perturbation screens, will increase the performance of models (de Boer and Taipale, 2024). To enable such breakthroughs, having the best possible benchmarks in the form of gold-standard data sets, and tools for mechanistic interpretation will be paramount.

Finally, having a strong community that promotes collaboration and communication will accelerate the pace by which progress is made. For this reason, we, as authors, are part of the Regulatory and Systems Genomics community of special interest (RegSys COSI), which organizes scientific sessions at the annual ISMB meeting. We encourage more participation from people of all backgrounds and career stages.

## Acknowledgments

We thank Anshul Kundaje, Žiga Avsec, Neşet Özel, Alireza Karbalayghareh and Christina Leslie for feedback on the manuscript. We also would like to thank all past organizers and participants of the RegSys COSI sessions at ISMB, which provided the basis for much of the work described in this perspective. Since the breadth of the topic is huge, there are many important contributions, and we apologize to those whose work we missed.

## Funding

J.Z. is funded by the Stowers Institute for Medical Research. F.A. is funded partially by NIH grant R35-GM128938. A.M is funded by the Research Council of Norway [187615], Helse Sør-Øst, and the University of Oslo through the Centre for Molecular Medicine Norway (NCMM) and the Norwegian Cancer Society [197884, 245890]. J.E. received funding from NIH grants DP1DA044371, U01MH130995, and U01HG012079. S.R. is funded partially by NIH grant R01-GM144708-03. S.S. is funded partially by NIH grant R35-GM131819. S.M. is funded by NIH grant R35GM144135 and the National Science Foundation grant CAREER 2045500. A.M.-R. is supported by Chan Zuckerberg Initiative Ancestry Network (2021-240438), CONACYT-FORDECYT- PRONACES grant numbers 11311 and 6390, and Programa de Apoyo a Proyectos de Investigación e InnovaciónTecnológica–Universidad Nacional Autónoma de México (PAPIIT-UNAM) grant number IN218023.

## Conflicts of Interest
The authors have no conflicts of interest to declare.

## Figure

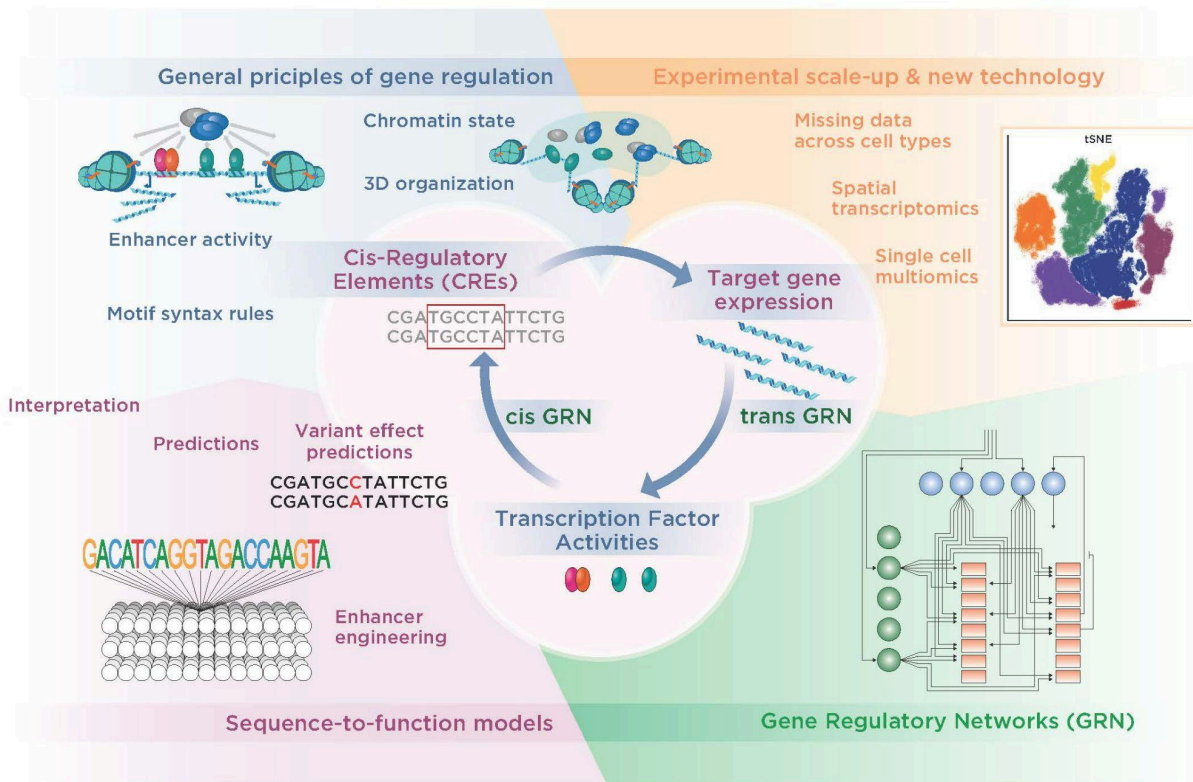

**Figure 1**. Regulatory genomics studies the intricate relationships between transcription factor activities and target gene expression, mediated by cis-regulatory elements. Researchers seek to identify general principles of gene regulation, such as how enhancer activities, motif syntax rules, 3D genome organization and chromatin states relate to each other and ultimately to gene expression. A major goal is to build accurate and generalizable sequence-function models that can not only reveal underlying mechanisms but make predictions of variant effects. Another major theme is the reconstruction of gene regulatory networks to describe biological processes of interest. Research into computational methods and models in regulatory genomics strives to make best use of diverse and rapidly advancing experimental technologies.